# Jamming, Friction and Unsteady Rheology

Mark O. Robbins
*Department of Physics and Astronomy, The Johns Hopkins University, Baltimore, MD 21218*
(March 24, 1999)

## I. INTRODUCTION

The very existence of friction is intimately related to jamming at the interface between two solids. Static friction is the force needed to overcome jamming and initiate sliding motion. Kinetic friction is the force that impedes sliding once it begins, and reflects energy dissipation due to sliding. In many cases the sliding is jerky due to periodic jamming and unjamming. This type of stick-slip motion can involve molecular scale or macroscopic slips between each jamming event. The intermittent nature of the motion can be important in determining how much energy is dissipated during sliding, and thus the kinetic friction.

In this overview chapter I will describe some of the many ways in which jamming produces friction on scales ranging from macroscopic to atomic. The goal will be to point to the commonality between processes in friction and the other manifestations of jamming discussed in this book. More complete treatments of friction can be found in a number of books [1–6], including the seminal book by Bowden and Tabor [1] and the collection of articles in "Fundamentals of Friction [2]."

The chapter begins with a brief review of the way friction is measured, the types of results that are observed, and what is known about the geometry of the contacts between two surfaces. Then the results of simple models of ideal, flat crystals are described. These reveal how the jamming that produces static friction is intimately related to the existence of metastable states, in this case due to the deformability of elastic solids. However, this type of metastability does not appear general enough to explain many examples of static friction.

Real surfaces are far from flat and chemically pure. Based on analogies to charge-density-wave conductors, flux lattices and other elastic systems, it seems natural that disorder such as surface roughness or chemical heterogeneity could lead to jamming. A brief review of studies of this effect is presented, and it is shown that pinning by disorder is exponentially weak in our three dimensional world.

A layer of some other material also normally separates two surfaces. This may be a thick lubricant film, dust, grease, wear debris produced by sliding, or a molecularly thin layer of hydrocarbon, water or other material adsorbed from the air. These additional objects between the two surfaces are called "third bodies" by tribologists and are known to affect the measured friction. Three of the papers following this overview chapter examine the behavior of a simple class of third bodies: Molecularly thin films of various fluids between ideal crystals. As the film thickness decreases to a few times a characteristic molecular diameter, these films undergo a jamming transition that is very similar to bulk glass transitions. Jammed systems exhibit a shear response that is typical of solid friction, and it is argued that jamming of third bodies provides a simple and general explanation for the prevalence of friction between the objects around us.

The chapter concludes by considering the unsteady stick-slip motion that often arises when systems become unjammed by sufficiently large stress. Similar unsteady motion occurs in all of the systems described in the following reprints [7–13]. Different types of stick-slip motion are identified, and some of its origins are explored.

## II. FRICTION MEASUREMENTS

The basic geometry of a friction measurement is shown in Figure 1. A solid is pushed down onto another solid with a normal load $L$. One then determines the relation between the lateral velocity $v$ and lateral force $F$, which is called the friction. In essence this is just a rheology measurement like those made on fluids, granular media, or foams. The difference is that one generally does not know the area, thickness or constituents of the interfacial region where shear occurs. Thus it is difficult to map $F$ and $v$ into the usual rheological variables: shear stress $\tau = F/A$ and shear rate $\dot\gamma \equiv \partial v_x/\partial z$. Progress has been made in developing well characterized systems, as discussed below. However, even in the best cases, one does not know how shear is distributed within the interfacial region.

Most macroscopic systems exhibit a force/velocity relation that is similar to the rheology of plastic materials (Fig. 2(a)) [1–3]. The velocity remains equal to zero until a threshold force called the static friction $F_s$ is exceeded. Once sliding has begun, a usually smaller kinetic friction force $F_k$ acts in the direction opposite to the motion. In many cases $F_k$ is relatively independent of velocity for reasons that will be discussed below.

One might expect a very different force/velocity relation for lubricated systems. If the lubricant was Newtonian and its thickness $h$ was constant, $F$ would rise

---

linearly with velocity, and there would be no static friction (Fig. 2(b)). However, if the velocity is kept at zero for a long enough interval, the applied load pushes any *fluid* lubricant out of the contacts. One then observes static friction and a velocity independent kinetic friction at very low velocities (Fig. 2(a)). As the velocity increases, hydrodynamic lift causes the lubricant layer to thicken, and the friction drops sharply. Eventually the lubricant thickness begins to saturate and the force rises. This force/velocity relation is called a Stribeck curve [14,15]. The large velocity region is well described by elasto-hydrodynamics [15,16], a continuum theory that includes the elastic deformation of the bounding solids as well as the hydrodynamics of the fluid between them. The small velocity region is very similar to what one finds for non-lubricated systems and presumably reflects the same jamming.

The two basic laws for static and kinetic friction that are still taught and used today were written down by Amontons exactly three hundred years ago and were known to da Vinci even earlier [4]. Amontons' first law is that friction is proportional to load, and the second is that friction is independent of the apparent macroscopic area of contact between the two solids. As a result the friction between two bodies is normally reported as a friction coefficient $\mu$ defined as the ratio of force to load: $\mu = F/L$. Different coefficients of friction are typically measured for static and kinetic friction, and denoted by $\mu_s$ and $\mu_k$, respectively. In some systems $\mu$ varies with load or surface area. We will see that these failures of Amontons' laws can help to understand their successes in many other systems.

As noted above, friction is a measure of rheology in the interfacial region, and it is not surprising that measured friction coefficients depend strongly on the material at the interface between the solids. To obtain reproducible results one must carefully control the surface conditions, including the ambient humidity. In some rare cases the two solids are in direct atomic contact. However, this is usually true only for experiments in ultra-high vacuum or when wear has exposed momentarily clean fracture surfaces. In most cases there are other objects between the two surfaces that are called "third bodies" by tribologists [17]. These range from sand grains, to dust, to wear debris, to thick lubricant films, down to single monolayers of grease, water or airborne hydrocarbons. In Amontons' original experiments a layer of grease was applied to the contacting surfaces. Most machines attempt to maintain a thick lubricant layer between the surfaces in order to minimize friction and wear. Theoretical treatments of friction have typically ignored third bodies, or used continuum mechanics to treat thick lubricant films. In part this may be due to the lack of precise information about the material at the interface. The ability to measure and control material in the contact is one of the great advances in experimental techniques over the last decade.

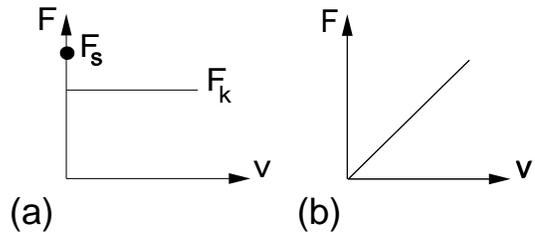

FIG. 2. Panel (a) shows the typical force/velocity relation for a friction measurement. There is no motion until the static friction $F_s$ is exceeded. Once motion starts, the kinetic friction $F_k$ resists motion. Panel(b) shows the force/velocity relation predicted for a Newtonian lubricant at constant film thickness or for sliding between two incommensurate solids.

### III. SIMPLE MODELS OF JAMMING

The existence of static friction implies that the surfaces and any material in the interfacial region must jam. The interfacial region becomes trapped in a local potential energy minimum. When an external force is applied to the top wall, it moves away from the minimum until the derivative of the potential energy balances the external force. The static friction is the maximum force the potential can resist, i.e. the maximum slope of the potential.

The simplest picture for jamming is that macroscopic peaks and valleys on the two surfaces interlock and prevent lateral motion as shown in Fig. 3(a). This was the basis of some of the oldest attempts to explain Amontons' laws. Parent and Euler [4] considered surfaces with no microscopic resistance to sliding, but roughened so that they make an angle $\theta$ relative to the average surface plane. A static friction $F_s = L \tan \theta$ must be overcome before the top surface can move up the ramp formed by the bottom surface and begin to slide. There is no dependence on surface area, and the predicted coefficient of friction $\mu_s = \tan \theta$ covers all possible values.

One might expect that Parent and Euler's model would apply to the system of glass spheres considered by Nasuno et al. in one of the following reprints [10]. Here glass spheres were glued to the bottom of a glass plate. The plate was then placed on a pack of spheres of the same size. One expects that the spheres on the plate will burrow between spheres in the pack when a load is applied. In order to move the plate, the spheres must be lifted out of the holes they created. The average angle required was worked out by Belidor in 1737 [4] and would give $\mu_s = 0.35$. However, glass beads can never slide past each other without microscopic friction as assumed by Parent, Euler and Belidor. In order to get a good model of friction in granular systems one must follow Coulomb [4] and include a microscopic friction coefficient as well as the mean slope. One then returns to the question of determining the origin of friction between two surfaces.

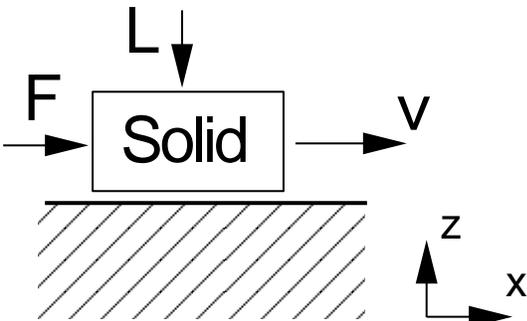

FIG. 1. Sketch of the typical geometry of a friction measurement. The relation between the tangential force $F$ and tangential velocity $v$ is measured as a function of the normal load $L$ that pushes the two surfaces together. The coordinate system used here is indicated.

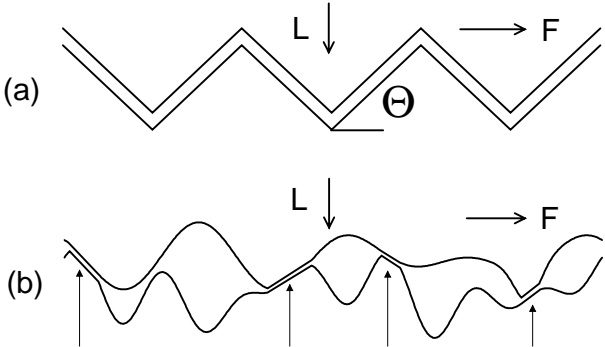

FIG. 3. (a) Ideal interlocking asperities making an angle $\theta$ relative to the average surface. (b) Contacts between random surfaces (arrows) occur at the peaks of isolated asperities. For every ramp that goes up there is another going down and the net force is zero. The vertical scale has been exaggerated and the angle in each contact would be orders of magnitude smaller [18].

Parent and Euler's picture of friction due to interlocking peaks, or asperities as they are more frequently called, can not explain many experimental trends and observations [1]. One is that decreasing $\theta$ by smoothing surfaces can actually increase static friction. Indeed magnetic hard disks are purposefully roughened to reduce $\mu_s$. Another is that experiments show that $\mu$ depends on the materials in contact more strongly than their surface morphology. For example, a single monolayer of fatty acid can reduce the friction by an order of magnitude without changing the surface roughness. Finally, Parent and Euler's model would give vanishing kinetic friction because the force applied to move up the ramps would be exactly balanced by the force exerted by the ramps on the way back down [19].

Over the last 50 years, a variety of experimental techniques [1,3,20,21] have shown that contacts between most macroscopic surfaces look like Fig. 3(b), where the roughness is exaggerated by magnifying the vertical scale. Molecules from opposing surfaces only contact where two peaks meet. The random spacing between peaks prevents the intimate mating envisioned in Fig. 3(a). Except for very soft materials, like rubber, the total area of intimate molecular contact, $A_{\rm real}$, is much less than the apparent geometrical area of the surfaces $A_{\rm app}$. Elastic deformation of the solids leads to relatively flat contact regions with small slopes and diameters of order one to ten micrometers. Parent and Euler's picture fails both because typical values of $\tan \theta$ are much smaller than $\mu_s$ and because the effect of the contacts averages to zero: For every contact with an upward slope there is another sloping down.

The implications of these experiments is that static friction must result from jamming within the areas of intimate molecular contact. The diameters of these contacts are much larger than molecular diameters (1 to 10$\mu$m vs. $\sim$1 nm), and are usually much larger than the thickness $h$ of the interfacial region where shear or jamming occurs. One thus expects that friction results from jamming within many independent regions of dimension $h$ within the contact. These regions will add constructively to the potential barrier that determines the static friction, leading to a friction that is linear in the real area of contact. Of course the local potential barriers will depend on the local pressure $p$ which will decrease with increasing $A_{real}$. We can summarize this by saying that a piece of contact area $dA$ contributes a static friction $dF_s = \tau_s(p)\, dA$ where $\tau_s$ is the yield stress at the local pressure.

The conclusions of the above paragraph might seem to conflict with Amontons' second law which says that the apparent area of contact does not influence the friction. However, the real and apparent areas are known to be very different. Bowden and Tabor [1] suggested a simple phenomenological theory for $\tau_s$ that not only yields Amontons' laws for a wide range of systems but can also explain experiments where Amontons' laws fail [7,22–25]. They assumed a simple linear relation between $\tau_s$ and $p$

$$\tau_s(p) = \tau_0 + \alpha p. \tag{3.1}$$

Then integrating the force over the entire contact area $A_{real}$ gives:

$$F_s = \tau_0 * A_{real} + \alpha * L, \tag{3.2}$$

where $L$ is the total normal load and the result is independent of the distribution of pressures in the contacts. Dividing by the total load gives the coefficient of static friction:

$$\mu_s = \alpha + \tau_0/\overline{p}, \tag{3.3}$$

where $\overline{p} = L/A_{\rm real}$ is the average pressure over all contacts. This expression for the coefficient of friction will be independent of load and macroscopic area if $\overline{p}$ is much larger than $\tau_0$ or if $\overline{p}$ is constant. As we now discuss, the latter holds for simple models of ideally plastic [1] or elastic [26,27] surfaces.

Plastic materials will deform to increase the contact area when the pressure exceeds the hardness of the material. Given the small values of $A_{\rm real}/A_{\rm app}$ observed in many systems, it seems likely that this failure condition may be met at typical values of the normal load [1,20]. The real area of contact at a given load will then grow until $p$ decreases to the hardness $H$, and the coefficient of friction will have the constant value:

$$\mu_s = \alpha + \tau_0/H. \tag{3.4}$$

Typical values of $\tau_0$ are usually small compared to $H$ and thus $\alpha$ will normally dominate the friction coefficient [28].

The argument for elastic surfaces is more complicated and the reader is referred to the original paper by Greenwood and Williamson [26] and a more recent paper by Volmer and Natterman [27]. The basic starting point is that real surfaces often exhibit power law noise spectra characteristic of self-affine fractals. The typical height variation $\delta h$ over a lateral distance $\delta l$ scales as $\delta h \sim \delta l^{0.5}$. The peaks on such surfaces have a broad distribution of heights. Increases in load will push the surfaces together, creating new contacts and expanding pre-existing contacts. The distribution of heights is such that the fraction of contacts that have a given area remains nearly unchanged, but the total number of each size increases linearly with load. Since the total area is proportional to load, the pressure remains constant. The constant value must be smaller than the hardness or plastic deformation will occur.

At large enough loads the real area of contact must become equal to the apparent area. Materials such as steel are so stiff and hard that the machines applying the load would be likely to fail before this condition was met. However, the rubber on tires and shoes is much more easily deformed. Experiments on polymers show that Amontons' laws fail when $A_{\rm real} = A_{\rm app}$, but the linear relation assumed by Bowden and Tabor (Eq. 3.1) is still followed [23]. The same linear relation holds for other situations where $A_{\rm real} = A_{\rm app}$, including experiments on $MoS_2$ and mica [7,22,24,25]. It even applies to the case of tape, where $P$ is zero, but a large force per unit area $\tau_0$ must be overcome to initiate sliding. Thus Eq. 3.1 is capable of describing a wide range of systems, encompassing both the failures and successes of Amontons' laws.

If static friction is just the rheology of a jammed system at the interface, one might expect that a relation like Eq. 3.1 would describe the yield stress of bulk systems. This expectation is born out by experiments on a variety of inorganic [29] and organic [30,31] solids that show a linear increase in bulk yield stress with hydrostatic pressure. The yield stress of granular materials is also known to increase with pressure, although this is normally attributed to the increased friction at contacts between grains due to the increased load [10,32–34]

To summarize this section, a wide range of experimental evidence indicates that static friction results from jamming of material at the interface between two solids. If the yield stress of this material increases linearly with pressure, then Amontons' laws and many exceptions to them can be understood. A yield stress that increases linearly with pressure is found in bulk experiments, but it is not clear what produces this type of jamming in friction.

## IV. CRYSTALLINE SURFACES IN DIRECT CONTACT

Theoretical studies of friction have historically focused on the case illustrated in Fig. 4(a) – two clean crystalline surfaces in direct contact [35–42]. The crystals are generally assumed to deform elastically, while sliding leads to plastic deformation at the interface. This implicitly assumes that the interactions across the interface are weaker than the interactions within the crystals, although the models are often extended beyond this regime.

Even the idealized case of Fig. 4(a) is difficult to analyze analytically. Figure 4 (b) and (c) show two even simpler one-dimensional (1D) models that have been used to understand whether and how two crystals can jam. Both models replace the bottom solid by a periodic potential, and retain only the bottom layer of atoms from the top wall. In the Tomlinson model (Fig. 4(b)), the remaining atoms are coupled to the center of mass of the top wall by springs, and coupling between neighboring atoms is ignored [35,36,43]. In the Frenkel-Kontorova model (Fig. 4(c)), the atoms are coupled to nearest-neighbors by springs, and the coupling to the atoms above is ignored [37–41]. Many hybrids and variants of these models have also been considered [42], but the qualitative features of the results are universal.

The dimensionless quantities that characterize both simple 1D models are the ratio between the lattice constants of the two surfaces, $\eta = b/a$, and the ratio of the strength of the periodic potential to that of the springs. If we assume that the force from the potential is a simple sine wave, $F_0 \sin(2\pi x/a)$, then we can characterize the potential's strength by the ratio $\lambda \equiv 2\pi F_0/ak$ of the maximum derivative of the force to $k$. The models are not self-consistent if $\lambda$ is much larger than one. In this limit the interactions at the interface are stronger than those in the bounding solids and the solids should yield before the interface. One should then consider models of sliding at an interface within the weaker solid.

The simplest case to consider is that of identical lattice constants, $\eta = 1$. In this case, all atoms go up and down over the potential in phase, just as in Fig. 3(a). The static friction per atom for both models is $F_s = F_0$, the maximum force from the periodic potential. The static friction is less than the maximum force in two or more dimensions because the atoms can move around the maxima in the potential and pass over a saddle point. This type of sliding occurs in internal shear of bulk fcc systems [44,45] and Hirano and Shinjo have noted this extra degree of freedom lowers the static friction relative to one-dimensional models [38].

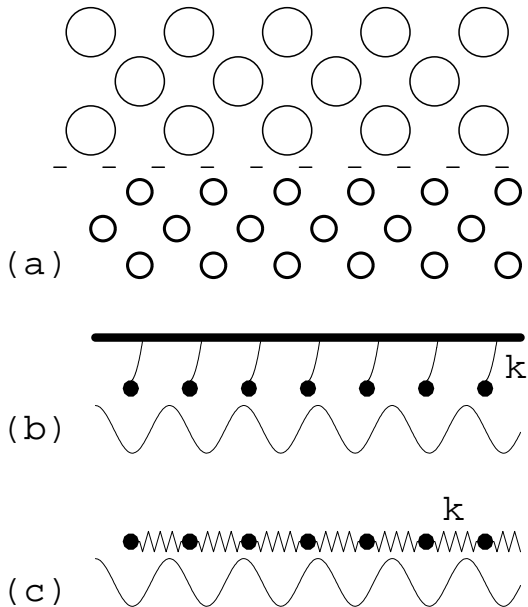

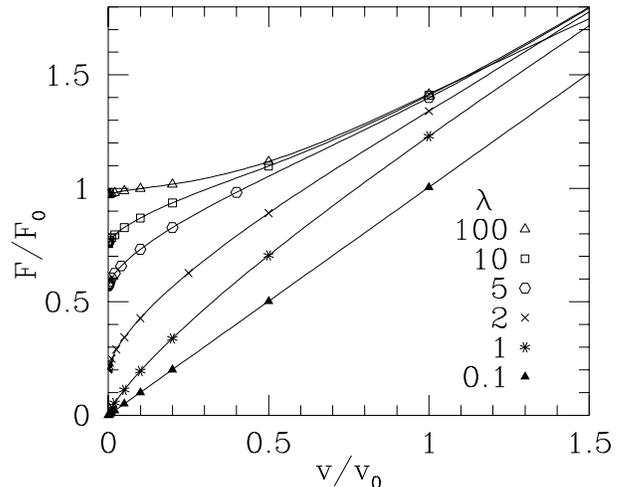

FIG. 5. Force vs. velocity for the Tomlinson model at the indicated values of $\lambda$. The force per atom $F$ is normalized by the static friction $F_0$, and the velocity is normalized by $v_0 \equiv F_0/\Gamma$ where $\Gamma$ is the phenomenological damping rate.

FIG. 4. (a) Two ideal, flat crystals making contact at the plane indicated by the dashed line. The nearest-neighbor spacings in the bottom and top walls are $a$ and $b$ respectively. The Tomlinson model (b) and Frenkel-Kontorova model (c) replace the bottom surface by a periodic potential $F_0 \sin(2\pi x/a)$. The former model keeps elastic forces between atoms on the top surface and the center of mass of the top wall, and the latter includes springs between neighbors in the top wall.

The kinetic friction is difficult to treat within these simple models because it necessarily involves dissipation. In the Tomlinson model each atom is an independent harmonic oscillator with no ability to dissipate. There are more degrees of freedom in the Frenkel-Kontorova model, but the vibrations are all exact normal modes with no allowed energy transfer. What is usually done in both models is to add a phenomenological viscous friction force, $F_{drag} = -\Gamma v$, to the equations of motion for each atom. This might be thought of as the damping associated with coupling to external degrees of freedom in a Langevin model, but the thermal noise term that would be coupled to the damping is usually ignored. In the following we specialize to the overdamped case where the inertia of the atoms can be ignored. The effects of inertia are discussed briefly in the section on stick-slip motion.

Although the above approximation involves assuming friction in order to calculate friction, one can still get non-trivial behavior. Indeed this version of the Tomlinson model is mathematically identical to simple models for Josephson junctions [46], to the single-particle model of charge-density wave depinning [47], and to the equations of motion for a contact line on a periodic surface [48,49]. Fig. 5 shows the results for the Tomlinson model as a function of the dimensionless strength of the interface potential, $\lambda$. When the potential is much weaker than the springs, the atoms can not deviate significantly from the center of mass. As a result, they go up and down over the periodic potential at constant velocity and the force averages out to zero. The kinetic friction is just due to the drag force on each atom and rises linearly with velocity. The same result holds for all spring constants in the Frenkel-Kontorova model with equal lattice constants.

As the potential becomes stronger, the periodic force begins to contribute to the kinetic friction of the Tomlinson model. There is a transition at $\lambda = 1$, and at larger $\lambda$ the kinetic friction remains finite in the limit of zero velocity. This constant approaches the static friction $F_0$ as $\lambda \to \infty$. The origin of this new low frequency behavior is the onset of multistability [36], a feature that is important in many jammed systems. The condition for metastability is that the spring and periodic forces balance:

$$k(x - x_{\rm CM}) = F_0 \sin[2\pi x/a], \qquad (4.1)$$

where $x_{\rm CM}$ is the center of mass position. As shown graphically in Fig. 6, there is only one solution for weak interfacial potentials ($\lambda < 1$), but there are an increasing number of metastable solutions as $\lambda$ increases beyond unity. Once an atom is in a given metastable minimum it is trapped there until the center of mass moves far enough away that the second derivative of the potential vanishes and the local minimum disappears. The equations of motion then become unstable, and the atom pops forward very rapidly to the next minimum. This only reduces the force from $F_0$ by about $ka$. Thus the friction force remains constant as $v$ goes to zero at a value that is of order the static friction and approaches it as $k$ goes to zero.

The above discussion explains the kinetic friction entirely in terms of the spring extension, but kinetic friction is fundamentally due to dissipation. The existence of a finite low velocity limit for the friction implies that the rate of dissipation rises linearly with velocity, rather than exhibiting the quadratic dependence characteristic of viscous fluids. If the rate of dissipation is linear in velocity there is the same amount of energy dissipated for every lattice constant advanced in the low velocity limit - how does this come about?

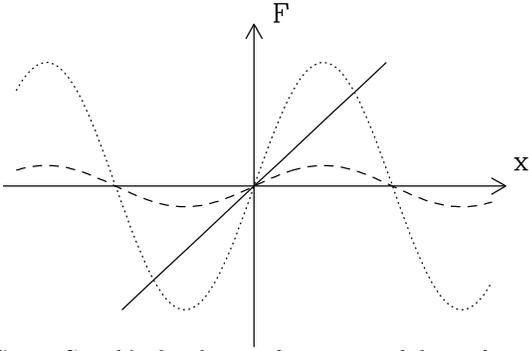

FIG. 6. Graphical solution for metastability of an atom in the Tomlinson model. The straight line shows the force from the spring for $x_{CM} = 0$. The dotted and dashed curves show periodic potentials with amplitudes corresponding to $\lambda = 3$ and 0.5, respectively. For $\lambda > 1$ there are multiple intersections with the spring force, indicating the existence of multiple metastable states.

The only mechanism of dissipation is through the phenomenological damping force, which is proportional to the velocity of the atom. The velocity is essentially zero except in the rapid pops that occur as a state becomes unstable and the atom pops to the next metastable state. During this pop the atom's velocity is always of order $v_0 \equiv F_0/\Gamma$ – independent of the average velocity of the center of mass. Moreover, the time of the pop is nearly independent of $v_{CM}$ and so the total energy dissipated is independent of $v_{CM}$. One can of course show that this dissipated energy is consistent with the limiting force determined from arguments based on the extension of the springs [48–50]. This basic idea that kinetic friction is due to dissipation during pops that remain rapid as $v_{CM} \to 0$ is very general, although the phenomenological damping used in the model is far from realistic.

The case of equal lattice constants is extremely unlikely, and becomes even more so when one considers three-dimensional objects with two-dimensional surfaces. Figure 7 shows that even identical crystals only match up when they are perfectly aligned. The most likely case is that the ratio of lattice constants is an irrational number. In this case the crystals are said to be incommensurate, while a rational ratio would correspond to a commensurate case. Incommensurate crystals share no common period, and thus lattice sites on the top wall lie at all possible positions relative to lattice sites on the bottom wall. This situation is necessarily translationally invariant, so there is no variation in the ground state with the position of the top wall.

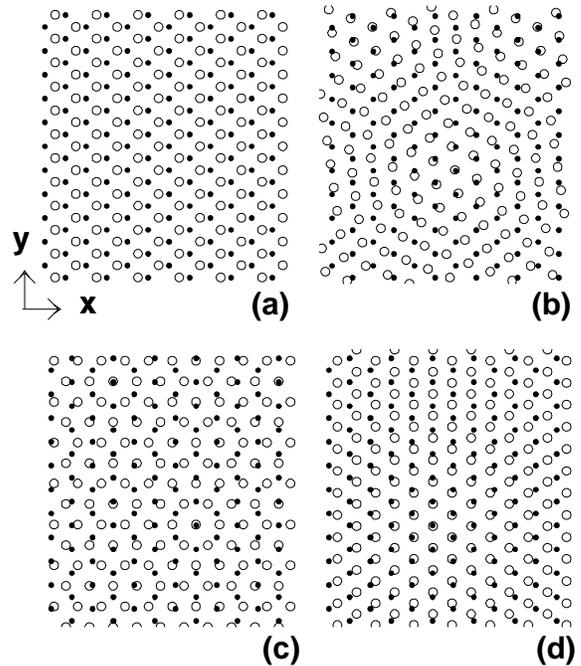

FIG. 7. The effect of crystalline alignment and lattice constant on commensurability is illustrated by projecting atoms from the bottom (filled circles) and top (open circles) surfaces into the plane of the walls. In (a)-(c) the two walls have the same structure and lattice constant but the top wall has been rotated by $0^o$, $8.2^o$ or $90^o$, respectively. In (d) the walls are aligned, but the lattice constant of the top wall has been reduced by a small amount. Only case (a) is commensurate. The other cases are incommensurate and atoms from the two walls sample all possible relative positions with equal probability. Reproduced by permission from Ref. [51].

One might assume that this translational invariance immediately implied that the static friction vanished for incommensurate systems. This is indeed the case in either the Tomlinson or Frenkel-Kontorova model if the potential is weak compared to the springs. In this case there is no multistability and atoms stay near their lattice sites. Since these sites sample all possible values of the force with equal probability, the average force, and hence the static friction, is zero. The situation changes when the interfacial potential becomes strong enough to introduce multistability for individual atoms. The Tomlinson case is the simplest. Although I have not seen it discussed for incommensurate systems, the mathematics is identical to Fisher's mean field theory of charge-density-wave pinning by random impurities [50]. If atoms start on their lattice sites and are then allowed to relax, they will reach exactly the same ground state energy for any center of mass position $x_{CM}$ of the top wall. All that changes is that the role of individual atoms is interchanged. If there is multistability, then some or all of the atoms will remain in these metastable states as $x_{CM}$ changes. Thus even though there is a solution for the new $x_{CM}$ with the ground state energy, atoms are unable to cross barriers to reach this solution. The result is a net force that reflects the metastable minimum that the system is trapped in. The static friction becomes non-zero when the potential is strong enough to produce multistability ($\lambda > 1$), and its value is just equal to the kinetic friction in the limit $v \to 0$ (Fig. 5).

The physics of the Frenkel-Kontorova model is much richer, but has the same qualitative features [37–41]. The main difference is that the static friction and ground state depend strongly on $\eta$. For any given irrational value of $\eta$ there is a threshold potential strength $\lambda_c$. For weaker potentials the static friction vanishes, and for stronger potentials metastability produces a finite static friction. The metastable states take the form of locally commensurate regions that are separated by domain walls where the two crystals are out of phase.

The one-dimensional Frenkel-Kontorova model has been analyzed in great detail [39–41]. Within the locally commensurate regions the ratio of the periods is a rational number $p/q$ that is close to $\eta$. In order to get a pinning force near $F_0$, the springs must be weak enough to allow a metastable state with $p = q = 1$. As $q$ increases, the pinning force decreases and the value of $\lambda_c$ rises. The range of $\eta$ where jamming occurs grows with increasing potential strength until it spans all values. At this point there is an infinite number of different metastable ground states that form a "Devil's staircase" as $\eta$ varies [39,40].

The two-dimensional case that would represent the interface between three dimensional crystals has received less attention, but seems to have the same general behavior [38,52,53]. The value of $\eta$ in such systems depends strongly on both the direction of sliding and the relative orientation of the crystalline axes of the two solids (Fig. 7). Thus, if this model described the metastability in real systems, one would expect strong variations in friction with alignment and sliding direction, including the absence of static friction in certain orientations [54]. Anisotropy is seen in certain systems [55–57], but the variations seen in typical experiments, where no effort is made to control orientation, are usually only of order 20%.

In order to get static friction at all orientations within the Frenkel- Kontorova model one would need to have a very strong interfacial potential compared to the cohesive potentials within each solid. As noted earlier, this is inconsistent with the assumption that sliding occurs at the interface rather than within the weaker solid. Strong interfacial interactions are most likely to occur between clean, reactive surfaces in ultrahigh vacuum. Experiments on clean metal surfaces do indeed show that the surfaces weld together at the interface and that sliding occurs through fracture within one of the solids. The friction forces are anomalously large, tend to depend on area, and material wears away at extraordinarily high rates.

In summary of this section, simple models predict that static friction between two flat crystals should only be observed in the unlikely case of commensurate walls, or when the walls are incommensurate and there is multistability. The condition for the latter is that the interfacial potential is large compared to interactions within the crystals. There have been relatively few experiments on flat, clean, crystalline surfaces, both because of the difficulty in making flat surfaces and because of the difficulty in removing chemical contamination. However, the experiments that have been done are consistent with the simple models described above.

One class of experiments measures the friction between an adsorbed layer and a substrate using a quartz crystal microbalance [58–60]. The substrate is atomically flat over large regions and the thickness of the adsorbed layer can be varied from a fraction of a monolayer up to several layers. When the layer is a fluid or incommensurate crystal there is no static friction. The friction varies linearly with velocity as in Fig. 2(b) and incommensurate crystals actually slide as much as an order of magnitude more easily than fluid layers of the same atom [58,60,61]. Because there is no metastability, the slope can be determined from the fluctuation-dissipation theorem. Theories based on dissipation through excitations of phonons [61–63] and electrons [64–66] have successfully explained the experimental results, although the relative importance of the two contributions has not been determined because of uncertainties in the interfacial potential [60].

Another experiment used a crystalline atomic-force-microscope tip that was rotated relative to a crystalline substrate. Static friction was observed when the crystals were aligned, and decreased below the threshold of detection when the crystals were rotated out of alignment [55]. Unfortunately the crystals were orders of magnitude smaller than typical contacts. Hirano and coworkers have also looked at the orientational dependence of the friction between macroscopic mica surfaces [57]. They found as much as an order of magnitude decrease in friction when the mica was rotated to become incommensurate. However, this large orientational anisotropy was only observed in vacuum. No anisotropy could be seen in ambient air.

A final set of experiments used $MoS_2$ lubricating two surfaces [67]. The $MoS_2$ forms plate-like crystals that slide over each other within the contact. They appear to be randomly oriented and thus incommensurate. In ultrahigh vacuum the measured friction coefficient was always lower than 0.002 and in many cases dropped below the experimental noise. In contrast, films exposed to air give friction coefficients of 0.01 to 0.05. This and the mica experiments provide a strong hint that contamination between surfaces is important to static friction as we discuss further below [51,54].

## V. ROUGHNESS AND CHEMICAL HETEROGENEITY

There are many examples of physical systems where disorder leads to multistability and a pinning force like static friction. The most studied examples are probably charge-density-wave conduction [47,50,68] and magnetic flux lattices [69,70], but other examples include fluid invasion of porous media [71–73], motion of magnetic domain walls [74–76], and spreading on disordered surfaces [48,49,77–79]. In each case, one finds that disorder produces pinning below a critical spatial dimension and no pinning in higher dimensions. The pinning is typically exponentially weak at the critical dimension.

To determine the critical dimension and pinning force one uses scaling theory and compares the energy gained by conforming to the disorder to the energy cost of the elastic deformation [80]. Spreading [77–79] and friction [27,81,82] are different than other systems because the elastic object that is deformed has a higher dimensionality than the disordered region. In spreading, the contact line where the fluid interface intersects the disordered solid is a 1D "surface" of a 2D elastic interface. In friction, the 2D surface of a 3D elastic solid interacts with a disordered substrate. In both cases an elastic deformation with a wavevector $q$ at the surface creates an elastic deformation over a distance of order $1/q$ into the bulk. The energy cost of deformations normally scales as $q^2$, but when integrated over the bulk, one factor of $q$ is canceled and the energy density scales as $q$. This changes the critical dimension from the charge-density wave case.

Consider the general case of a $d$ dimensional system with a $d-1$ dimensional surface that interacts with a random potential. The energy per unit surface area needed to deform a region of diameter $l$ scales as $q \sim 1/l$. The energy density gained by conforming to the disorder in this region scales as $1/l^{(d-1)/2}$ if the disorder is uncorrelated. This is just the usual result that the fluctuation in the mean of $n$ independent quantities scales as $1/\sqrt{n}$ with $n$ proportional to the area of the deformed region. Comparing the elastic cost to the gain one finds that disorder always wins at lengths longer than some size $l_p$ for $d < 3$, but not for $d > 3$. Thus contact lines are always pinned, and friction is a marginal case.

Volmer and Natterman [27] have considered the case of friction in detail. There is always pinning, even for weak disorder, but the pinning force is exponentially small. By making several assumptions one can obtain a static friction that satisfies Amontons' laws. However, it seems unlikely that these assumptions would be satisfied broadly enough to provide a general explanation of Amontons' laws. For example, one key assumption is that $l_p$ is comparable to the diameter of individual contacts. Given the large size of contacts (10 $\mu$m), this would require very weak disorder. One can estimate $\tau_s$ using the fact that the pinning and elastic energies are roughly equal at $l_p$. The elastic energy is the cost of deforming by a lattice constant $a$ over a wavelength of $l_p$, while the theoretical yield stress $\tau_y$ is the stress needed to deform by $a$ over a wavelength of a lattice constant. Thus $\tau_s \sim \tau_y l_p/a \sim 10^{-4} \tau_y$ which is far smaller than experimental values. In the case of plastic materials it would imply $\mu < 10^{-4}$.

The basic reason that it is difficult for incommensurate potentials or disorder to pin crystals is that the atoms must move by distances of order a lattice constant in order to produce a metastable state. This type of plastic deformation can only occur when the forces are strong enough to tear the crystal apart. Stresses large enough to produce plastic deformation may occur in some cases, but are expected to lead to extremely high wear rates that would not be acceptable in practical applications. Replacing the crystals with elastic amorphous walls, does not resolve this problem. However, as we now discuss, introducing easily deformable "third bodies" between the solid walls allows them to jam together without producing excessive wear.

## VI. JAMMING IN CONFINED FLUIDS

The Surface Force Apparatus (SFA) allows the mechanical properties of fluid films to be studied as a function of thickness over a range from hundreds of nanometers down to contact. The fluid is confined between two atomically flat surfaces. The most commonly used surfaces are mica, but silica, polymers, and mica coated with amorphous carbon, sapphire, or aluminum oxide have also been used. The surfaces are pressed together with a constant normal load and the separation between them is measured using optical interferometry. The fluid can then be sheared by translating one surface at a constant tangential velocity [7,83] or by oscillating it at a controlled amplitude and frequency in a tangential [8,84,85] or normal [86,87] direction. The steady sliding mode mimics a typical macroscopic friction measurement, while the oscillatory mode is more typical of bulk rheological measurements. Both modes reveal the same sequence of transitions in the behavior of thin films.

When the film thickness $h$ is sufficiently thick, one observes the rheological behavior typical of bulk fluids [86,87]. Flow can be described by the bulk viscosity $\mu_B$ with the usual no-slip boundary condition - equal velocities of the fluid and bounding solid at their interface. The shear stress $\tau$ is just

$$\tau = \mu_B v/h \qquad (6.1)$$

where $v$ is the velocity difference between the two walls.

As the film thickness decreases below 100nm, the behavior begins to change [86,87]. Experiments can not determine whether this is due to changes in viscosity or in boundary condition, because the flow profile is not measured. However simulations indicate that changes in viscosity do not become significant until the thickness becomes much smaller [9,88–92]. The experiments are consistent with small deviations from the no-slip boundary condition that are also found in simulations. These change the effective width of the film from $h$ to $h + 2S$ in Eq. 6.1 where $S$ is called the slip length and the factor of two comes from the presence of two interfaces. The slip length quantifies the degree to which the viscosity near the interfaces is different from the bulk value. In some cases the fluid slips more easily over the wall and $S$ is positive. In other cases a layer of fluid solidifies on to the wall and $S$ is negative. The displacement of the slip plane is comparable to the diameter of fluid molecules in most experiments, and several simulation studies have examined the factors that determine this boundary condition [90,92–96].

When the film thickness decreases below of order ten molecular diameters, the deviations from bulk behavior become dramatic [7,85]. The shear stress becomes far too large to interpret in terms of a bulk viscosity and slip length, because the width of the fluid region $h + 2S$ would have to be much smaller than an Ångström. Experimental results are often expressed in terms of Eq. 6.1 with an effective viscosity $\mu_{eff} \equiv \tau h/v$ and effective shear rate $\dot{\gamma}_{eff} \equiv v/h$. The following reprints by Gee et al. [7] and Demirel and Granick [8] show that low velocity values of $\mu_{eff}$ rise many orders of magnitude beyond $\mu_B$ and that relaxation times determined from the shear rate dependence of $\mu_{eff}$ increase even more dramatically over bulk values. Both changes are strong evidence that the film is becoming jammed. By the time $h$ reaches two or three molecular diameters, most films behave like solids.

In some cases the transition from liquid to solid behavior of the film appears to occur discontinuously, as if the film underwent a first-order freezing transition [97]. Simulations suggest that this is most likely to happen when the crystalline phase of the film is commensurate with the solid walls, and when the molecules have a relatively simple structure that facilitates order [7,9,98–100]. In most cases the transition looks like a continuous glass transition, but at temperatures and pressures that are far from the glass region in the bulk phase diagram. In fact, the molecules may not readily form glasses in the bulk. This suggests that one must treat the film thickness as an additional thermodynamic variable that may shift or alter phase boundaries. Cases where a single interface stabilizes a different phase than the bulk are central to the field of wetting. The presence of two interfaces separated by only a few nanometers leads to more pervasive phase changes [7].

In one of the following reprints [7], Gee et al. describe the "liquid to solid-like transitions" of five different liquids, cyclohexane, octamethylcyclotetrasiloxane (OMCTS), $n$-octane, $n$-tetradecane, and a branched isoparaffin 2-methyloctadecane. The liquids have different molecular structures that lead to different types of ordering in thin layers, and provide a reasonable sampling of the types of fluids that have been studied. For simple molecules one finds strong oscillations in the normal force on the walls as the film thickness decreases [89,92,101,102]. The period of the oscillations is a characteristic molecular dimension, and the oscillations reflect ordering of molecules into layers that are parallel to the bounding walls. The factors that control layering are discussed in detail in the reprint by Thompson et al. [9], references therein, and in some subsequent work [99].

For the five liquids studied by Gee et al., and many others studied later [25,103–106], the viscosity and relaxation times become too large to measure when the thickness has dropped to one to three molecular diameters. These thin films are jammed and resist shear like a solid. No motion occurs until a yield stress is exceeded, and the calculated yield stress is comparable to values for bulk solid phases of the molecules at higher pressure or lower temperature. The response of the walls to a lateral force becomes typical of static friction, and Gee et al. analyze their results in terms of equation 3.2 using $S_c$ and $C$ in place of $\tau_0$ and $\alpha$, respectively. The values of $\tau_0$ are of order a few to 20MPa, and values of $\alpha$ range from 0.3 to 1.5. However, the latter are influenced by the change in film thickness at the relatively low pressures used ($< 40$ MPa) and would probably not be representative of the behavior in a typical mechanical device. Gee et al. also describe unsteady stick-slip motion of the walls for certain fluids [7,107]. This is discussed in the next section.

In another reprint, Demirel and Granick [8] examine the transition from liquid to solid in OMCTS, following an analogy to studies of bulk glass transitions [108,109]. They measure the frequency dependence of the real and imaginary parts of the elastic moduli of thin films. They find that all the data can be collapsed onto a universal curve using a generalization of time-temperature scaling [108,109], and conclude by comparing the spectrum of relaxation times to those of bulk glasses.

Figure 8 shows that simulation results for the viscosity of thin films can also be collapsed using Demirel and Granick's approach [110,111]. The simulation model used is described in detail in the reprint by Thompson et al. [9]. Data for different thicknesses, normal pressures, and interaction parameters taken from Figs. 9, 15 and 16 of this paper were scaled by the low shear rate viscosity $\mu_0$. The shear rate was then scaled by the rate $\dot{\gamma}_c$ at which the viscosity dropped to $\mu_0/2$. Also shown on the plot (circles) are data for different temperatures that were obtained for longer chains in films that are thick enough to exhibit bulk behavior [110–112]. The data fit well on to the same curve, providing a strong indication that a similar glass transition occurs whether thickness, normal pressure, or temperature is varied. In all cases the high shear rate data is characterized by a power law shear thinning with an exponent near -2/3. Similar exponents are observed in experimentally measured viscosities [105].

As in the case of bulk glasses, the issue of whether there is a true glass transition at finite temperature is controversial. The simulations seem to show a clear change in behavior, and the divergence of the relaxation time $\dot{\gamma}_c^{-1}$ and $\mu_0$ can be fit to a free volume theory [9,111,113].

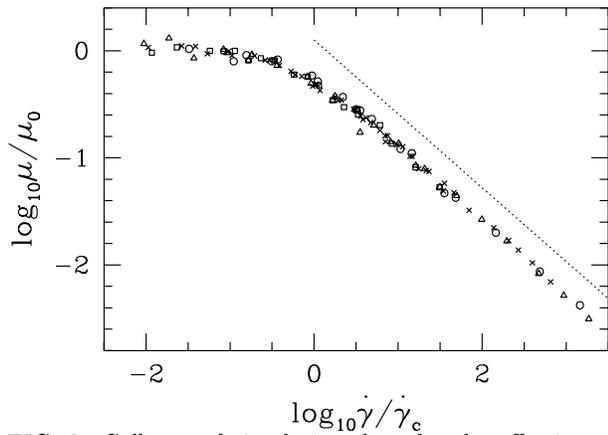

FIG. 8. Collapse of simulation data for the effective viscosity vs. shear rate as a glass transition is approached by decreasing temperature (circles), increasing normal pressure at fixed number of fluid layers (triangles) or decreasing film thickness at fixed pressure with two different sets of interaction potentials (squares and crosses). The dashed line has a slope of -0.69. Reproduced by permission from Ref. [111].

In the glassy state the response shows a constant yield stress which would correspond to a slope of -1 on Fig. 8. There is no case where we see a smooth crossover to this slope from the value of -0.69 indicated by the dashed line in Fig. 8, but our data is in a very different frequency range than experiments and spans fewer decades. Demirel and Granick get a reasonable collapse of data at all thicknesses [8]. This would indicate that the relaxation time never truly diverges. However, there is some ambiguity in their collapse of data for 1 and 2 layers, since the moduli are nearly independent of rate at these thicknesses.

Whether the relaxation time and viscosity truly diverge or not, they clearly become large compared to measurement times. Thin films of some molecules support stress over at least several days in the SFA [7,24]. The pressures in most macroscopic friction measurements are roughly two orders of magnitude higher, making the relaxation times even longer. Thus jamming of thin fluid layers between asperities provides a possible explanation for the prevalence of static friction in macroscopic experiments. This has been explored in a recent paper [51] using the wall geometries shown in Fig. 7. The shear stress was found to follow Bowden and Tabor's phenomenological law (Eq. 3.1) up to the largest pressures studied ($\sim 1.5$ GPa). In contrast to the case of bare walls, where the friction can depend strongly on the degree of incommensurability [37–41], $\tau$ was nearly independent of the orientation of the walls and the direction of sliding. The results were also independent of other factors that are not controlled in the experiments, such as the length of the molecules and their density.

## VII. STICK-SLIP MOTION

Any jammed system can be unjammed by the application of a sufficiently large shear stress. The subsequent dynamics depend on many factors, including the types of metastable state in the system, the times needed to transform between states, and the mechanical properties of the device that imposes the stress. At high rates or stresses, systems usually slide smoothly. At low rates the motion often becomes intermittent, with the system alternately jamming and unjamming. Everyday examples of such stick-slip motion include the squeak of hinges and the music of violins.

The alternation between jammed and unjammed states of the system reflects changes in the way energy is stored. While the system is jammed, elastic energy is pumped in to the system by the driving device. When the system unjams, this elastic energy is released into kinetic energy, and eventually dissipated as heat. The system then jams once more, begins to store elastic energy, and the process continues. Both elastic and kinetic energy can be stored in all the mechanical elements that drive the system. The whole coupled assembly must be included in any analysis of the dynamics.

The simplest type of intermittent motion is illustrated by the multistable regime ($\lambda > 1$) of the Tomlinson model (Fig. 4(b)). Consider an atom that starts at rest in a metastable state and is pulled by a wall moving with constant center of mass velocity $v$. As the wall moves forward, the force from the spring increases. The atom moves gradually up the local metastable potential well, storing more and more energy as the force increases. When the local well becomes unstable, the atom pops rapidly to the next potential well. In the process, potential energy is converted to kinetic energy and dissipated by the damping term. As noted in Section IV, this dissipation approaches a constant value as $v_{\rm CM} \to 0$ because the characteristic atomic velocity during each pop remains large in this limit ($v_{max} \sim v_0 = F_0/\Gamma$). As $v_{\rm CM}$ increases, the motion becomes smoother and smoother, and for $v_{\rm CM} > v_0$ there is no period where the atoms are jammed.

This type of intermittent motion is very similar to the atomic-scale stick-slip observed with atomic force microscopes (AFMs) [114]. Even though the tip usually contains several atoms, its potential energy varies with the periodicity of the underlying surface. The effective spring constant reflects the entire mechanical system that imposes stress, including the internal stiffness of the tip, the cantilever that moves it, and the piezoelectric devices that displace the cantilever. In most cases, $k$ is nearly equal to the cantilever stiffness, and the force is measured from the cantilever deflection. As the tip scans over the surface one often sees oscillations of the force that reflect sticking and slipping with the periodicity of an atomic spacing [114,115]. This indicates that $\lambda > 1$ and an estimate of $\lambda$ can be obtained by reversing the direction of the scan and seeing how far one must move before popping in the opposite direction.

The above examples of stick-slip motion involve a simple ratcheting over a surface potential through a regular series of hops between neighboring metastable states. The slip distance is determined entirely by the periodicity of the surface potential. The granular media [10,12] and foams [11] discussed in the following reprints have a much richer potential energy landscape due to their many internal degrees of freedom. As a result, stick-slip motion between neighboring metastable states can involve a complicated sequence of slips of varying length. For example, Nasuno et al.'s study of stick-slip in bead packs [10] reveals an erratic series of slip distances that are much smaller than the bead diameter at low velocity (Fig. 2(b) of Ref. [10]). This implies that their system has metastable states that are separated by very small wall displacements. Presumably a small cluster of beads becomes unstable and rearranges at each slip event. This shifts stress to the remaining beads, causing them to move slightly within their potential wells. The net displacement will be much less than a bead diameter if the fraction of the beads involved in the rearrangement is small. Such local rearrangements are studied in the reprints by Veje et al. [12] and Gopal and Durian [11] who consider sand and foam, respectively.

Many examples of stick-slip involve a rather different type of motion that can lead to intermittency and chaos [116,117]. Instead of jumping between neighboring metastable states, the system slips for very long distances before sticking. For example, Gee et al. [7] observe slip distances of many microns in their studies of confined films. This distance is much larger than any characteristic periodicity in the potential, and varies with velocity, load, and the mass and stiffness of the SFA. The fact that the SFA does not jam after moving by a lattice constant indicates that sliding has changed the state of the system in some manner, so that it can continue sliding even at forces less than the yield stress.

One simple property that depends on past history is the amount of stored kinetic energy. This can provide enough inertia to carry a system over potential energy barriers even when the stress is below the yield stress. Inertia is readily included in the Tomlinson model and has been thoroughly studied in the mathematically equivalent case of an underdamped Josephson junction [46]. One finds a hysteretic response function like that sketched in Fig. 9. Both static and moving steady-states exist over a range of forces between $F_{\min}$ and the static friction $F_s$. Inertia leads to similar hysteresis in the dynamics of a sand grain along a sand pile [118].

Hysteresis in the Tomlinson model can produce stick-slip motion if the top wall has its own mass $M$ and is connected through a much weaker external spring $k'$ [119] to a stage that moves at a constant velocity $u$ less than $u_{\min}$ (Fig. 9). The time-dependence looks much like that in the more complicated system shown in Fig. 10. While the wall is pinned, the force exerted by the external spring $k'$ rises at a constant rate $dF/dt = k'u$. When the force exceeds the static friction, the wall starts to move at a velocity greater than $u_{\min}$. As it catches up with the moving stage, the force drops. When it falls to $F_{\min}$, the wall stops, and the process repeats.

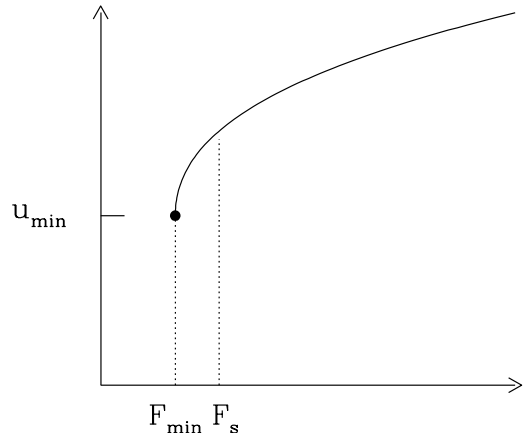

FIG. 9. Sketch of the force-velocity curve (solid line) for the underdamped Tomlinson model. The system is bistable over the range of forces between $F_{\min}$ and the static friction $F_s$ (dotted lines). The size of this region depends on the degree of underdamping. Sliding velocities less than $u_{\min}$ are unstable.

The systems studied in the following reprints have many additional structural degrees of freedom that can change with time to produce hysteresis. Such changes lead to complex multivalued force/velocity curves [20,117,120,121] like that in Fig. 3(b) of Ref. [10] that are often modelled with "rate-state" equations. The changes in the system with sliding or sticking time are lumped into a single phenomenological state variable whose value depends on the sliding history [20,117,120,121]. Unfortunately it is difficult to determine the nature of the state variable without making microscopic structural measurements of the system. This can be difficult in experiments, but is relatively easily done in simulations.

One structural variable that is found to change in both experiments and simulations is the density. The degree of dilation is known to play a critical role in the yield and dynamics of granular media [122], and its role is discussed in the following reprints by Nasuno et al. [10] and Veje et al. [12]. Simulations of stick-slip motion in confined fluid films (Fig. 10) also showed dilation during slip [107], and dilation was subsequently measured in experiments on these systems [126]. In all cases, work must be done to expand the volume against the imposed external pressure (or gravity). Once this is done there is more room for shear to occur. The system may be able to keep sliding as long as it takes more time for the volume to contract, than for the system to traverse between metastable states.

Simulations of stick-slip in confined films show that other types of structural change can occur. In cases where confinement induces a crystalline structure in the film, the film may undergo periodic melting and crystallization transitions as it goes from static to sliding states and then back [107,124]. An example of this behavior is shown in Fig. 10. Glassy films [107,111] and sand [125] can undergo periodic "melting" and jamming transitions. As in equilibrium, the structural differences between glass and fluid states are small. However, there are strong changes in the self-diffusion and other dynamic properties when the film goes from the static glassy to sliding fluid state. These melting and freezing transitions are induced by shear and not by the negligible changes in temperature. Shear-melting transitions at constant temperature have been observed in both theoretical and experimental studies of bulk colloidal systems [44,45].

In the cases just described, the entire film transforms to a new state, and shear occurs throughout the film. Another type of behavior is also observed. In some systems shear is confined to a single plane - either a wall/film interface, or a plane within the film [110,111]. There is always some dilation at the shear plane to facilitate sliding. In some cases there is also in-plane ordering of the film to enable it to slide more easily over the wall [110,111]. This ordering remains after sliding stops, and provides a mechanism for the long-term memory seen in some experiments [7,103,123].

The dynamics of the jamming and unjamming transitions during stick-slip motion are crucial in determining the range of velocities where it is observed, the shape of the stick-slip events, and whether stick-slip disappears in a continuous or discontinuous manner [6,104,107,116,121]. Current models are limited to energy balance arguments [107] or phenomenological models of the nucleation and growth of jammed regions [6,104,121]. Microscopic models and detailed experimental data on the jamming and unjamming process are lacking. Kinetic models for other systems, like that in the reprint by Farr et al. [13], may help to understand these complex processes.

## VIII. SUMMARY AND CONCLUSIONS

The preceding sections have described the key role jamming plays in many aspects of friction. At the same time they have attempted to draw parallels to other manifestations of jamming in this book.

The ubiquitous observation of static friction implies that jamming occurs at the interfaces between almost all objects. Jamming of crystals, pinning by disorder, and jamming of third bodies between the surfaces were discussed. The latter mechanism seems to provide the most robust explanation of static friction and Amontons' laws [51,54]. Third bodies are known to be present on almost all interfaces, whether in the form of air born hydrocarbons, wear debris or dust. Their interactions are dominated by hard core repulsions at typical pressures, and one may expect progress in the understanding of jamming in granular media, bulk glasses, and other systems to be directly relevant to our understanding of friction.

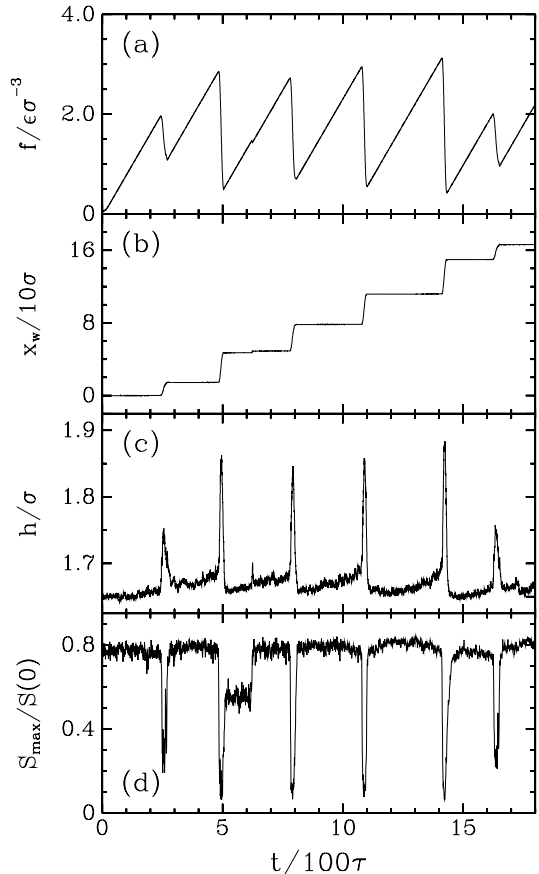

FIG. 10. Time profiles of the (a) frictional force per unit area $f$, (b) displacement of the top wall $x_w$, (c) wall spacing $h$, and (d) Debye-Waller factor $S_{max}/S(0)$ during stick-slip motion of spherical molecules that form two crystalline layers in the static state. Note that the system dilates (c) during each slip event. The coinciding drop in Debye-Waller factor shows a dramatic decrease from a typical crystalline value to a characteristic value for a fluid. Quantities are normalized by $\epsilon$, $\sigma$ and $\tau$ which are the characteristic energy, length and time scales of the Lennard-Jones potential.

Two types of transition to a jammed state were described. The first is the solidification of fluid films as the thickness decreases. Both experiment and theory indicate that this transition is closely related to equilibrium glass transitions in many systems [7–9,111], although crystallization may also occur [97,100,107]. The second type of transition occurs during stick-slip motion where the system alternates between jammed and unjammed states. These non-equilibrium transitions present an even greater theoretical challenge, because of the extra dynamic degrees of freedom and the lack of general principles for determining the stable state of non-equilibrium systems.

## ACKNOWLEDGMENTS

I thank my collaborators, A. R. C. Baljon, M. Cieplak G. S. Grest, G. He, R. Mountain, M. Müser, E. D. Smith, and P. A. Thomspon, for their help and insight. I would also like to thank the many experimentalists, especially S. Granick, J. Klein, J. Krim, J. N. Israelacvhili and P. M. McGuiggan, whose new techniques have changed the types of questions we can ask about friction. Finally, I gratefully acknowledge support from the National Science Foundation through Grant No. DMR-9634131 and through the Institute for Theoretical Physics.